\documentclass[aps,prb,twocolumn,groupedaddress]{revtex4}

\usepackage{amsfonts}
\usepackage{graphicx}
\begin{document}

\title{Low Frequency Transport Measurements in GdSr$_2$RuCu$_2$O$_8$}

\author{A. Vecchione}
\affiliation{Departement of Physics \lq\lq E. R. Caianiello\rq\rq
, and INFM Research Unit, University of Salerno, via
S. Allende, I-84081 Baronissi, Italy. }
\author{D. Zola}
\affiliation{Departement of Physics \lq\lq E. R. Caianiello\rq\rq
, and INFM Research Unit, University of Salerno, via
S. Allende, I-84081 Baronissi, Italy. }

\author{G. Carapella}
\thanks{Corresponding author}
\email[\newline e-mail: ]{giocar@sa.infn.it}
\thanks{\newline FAX: +3908965390 \\}
\affiliation{Departement of Physics \lq\lq E. R. Caianiello\rq\rq
, and INFM Research Unit, University of Salerno, via
S. Allende, I-84081 Baronissi, Italy. }

\author{M. Gombos}
\affiliation{Departement of Physics \lq\lq E. R. Caianiello\rq\rq
, and INFM Research Unit, University of Salerno, via
S. Allende, I-84081 Baronissi, Italy. }

\author{S. Pace}
\affiliation{Departement of Physics \lq\lq E. R. Caianiello\rq\rq
, and INFM Research Unit, University of Salerno, via
S. Allende, I-84081 Baronissi, Italy. }

\author{G. Costabile}
\affiliation{Departement of Physics \lq\lq E. R. Caianiello\rq\rq
, and INFM Research Unit, University of Salerno, via
S. Allende, I-84081 Baronissi, Italy. }

\author{C. Noce}
\affiliation{Departement of Physics \lq\lq E. R. Caianiello\rq\rq
, and INFM Research Unit, University of Salerno, via
S. Allende, I-84081 Baronissi, Italy. }

\date{\today}

\begin{abstract}
Low frequency transport measurements are performed on
GdSr$_2$RuCu$_2$O$_8$ pellets. The observed current-voltage curves
are qualitatively explained in the framework of a simple
phenomenological model accounting for  coexistence of
ferromagnetism and superconductivity in the sample. A Curie
temperature $T_{cM}$=133~K and a superconducting critical
temperature $T_{cS}$=18~K, with an onset temperature
$T_{cO}$=33~K, are extracted from the analysis of the
current-voltage curves.
\end{abstract}

\pacs{74.50.+r}

\maketitle

%\begin{multicols}{2}
%\narrowtext

\section{Introduction}

The interplay of magnetism and superconductivity is a fundamental
problem in condensed-matter physics and it has been studied
experimentally and theoretically for almost four decades. These
two cooperative phenomena are mutually antagonists. Indeed, the
superconductivity is associated with the pairing of electrons
states related to time reversal while in the magnetic states the
time-reversal symmetry is lost and therefore there is a strong
competition with superconductivity\cite{maple}. However, Schlabitz
{\it et al.}\cite{schlabitz} showed that surprisingly magnetism
and superconductivity could coexist in the heavy fermion compound
URu$_2$Si$_2$. Other heavy fermion superconductors have also been
shown to exhibit magnetic moments in their superconducting phase
\cite{review}. All these compounds contain rare-earth or actinide
ions with very localized 4f or 5f orbitals, strongly interacting
with the conduction band electrons. This is in contrast to the
Chevrel phases where magnetism and superconductivity coexist
because the magnetic moments responsible for magnetism are only
very weakly coupled with the electrons that form the condensate
\cite{fisher}.

Nevertheless, there are been a number of recent studies reporting
the coexistence of superconductivity and magnetic order in
R$_{1.4}$Ce$_{0.6}$Sr$_2$RuCu$_2$O$_{10-\delta}$ \cite{felner1}
and RSr$_2$RuCu$_2$O$_8$
\cite{tallon,tallon2,williams} where R=Gd, Eu.
These latter compounds were originally synthesized by Bauernfeind
{\it et al.} \cite{bauer} and Felner and co-workers \cite{felner}.
Most recent reports have focused on GdSr$_2$RuCu$_2$O$_8$, which
has a unit cell similar to that of the YBa$_2$Cu$_3$O$_7$ high
temperature cuprate, where there are two CuO$_2$ layers and one
RuO$_2$ layer with the CuO$_2$ and RuO$_2$ layers being separated
by insulating layers. Magnetization and muon spin rotation
studies \cite{tallon2} have shown that there exists a magnetic
ordering transition at temperature much greater than the
superconducting transition temperature. Some studies have been
interpreted in terms of ferromagnetic order arising from the Ru
moment in the RuO$_2$ layers. This idea has generated considerable
interest because ferromagnetic order and superconductivity are
mutually competing processes and could only coexist via some
accommodation of the respective order parameters by a spatial
modulation\cite{chu} or via the formation of a spontaneous vortex
phase \cite{vortex}. However, powder neutron diffraction study
\cite{neutron} has shown that while there is a small ferromagnetic
component, the low-field magnetic order is predominantly
antiferromagnetic. These contrasting reports cast some doubt about
the magnetic nature of this compound and at the present the
situation has not been completely clarified. The aim of this paper
is to give a contribution to this discussion. Indeed, we have
found that transport measurements performed on
GdSr$_2$RuCu$_2$O$_8$ sample are in agreement with predictions of
a simple phenomenological model where {\it ferromagnetism} and
superconductivity coexist. From the experimental results a Curie
temperature $T_{cM}$=133K and a superconducting critical
temperature $T_{cS}$=18K, with an onset temperature $T_{cO}$=33K,
have been inferred.

\noindent
The paper is organized as follows. In Section II the
sample preparation is discussed. A phenomenological model for
expected current-voltage curves is then given in Section III.
Experimental results are presented and discussed in connection
with the theoretical prediction of the proposed  model in Section
IV.  Some conclusions are finally given in the last
 Section.

\section{Sample preparation and characterization}
\begin{figure}[!h]
\includegraphics[width=7cm]{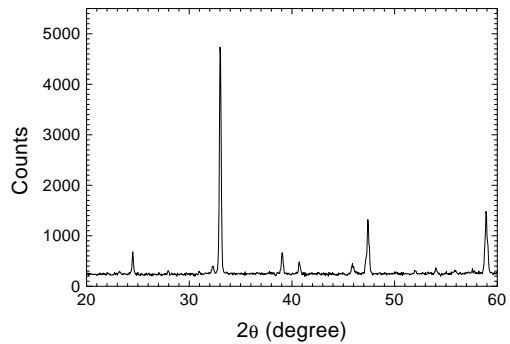}
\caption{(a) X-ray diffraction patterns of the
GdSr$_2$RuCu$_2$O$_8$.} \label{fig1}
\end{figure}
Precursors powders have been synthesized starting from the pure
binary oxide and carbonate powders, Gd$_2$O$_3$, SrCO$_3$, CuO,
and RuO$_2$, mixed together in the proper amount and solid state
reacted. The mixed powder was calcinated in air at 960 $^\circ$C
for 10~h. Annealing in flowing pure nitrogen at 1000~$^\circ$C
during 10~h was performed to reduce the formation of undesired
phases such as SrRuO$_3$ \cite{bauer}.
\begin{figure}[tbp]
\includegraphics[width=7cm]{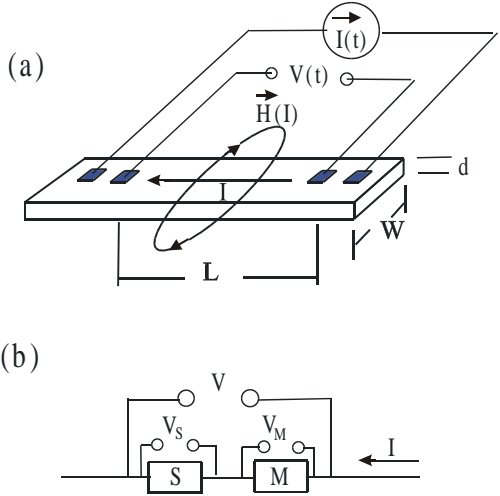}
\caption{(a) Sketch of wire connections used to measure the
GdSr$_{2}$RuCu$_{2}$O$_{8}$ pellet. (b) Measured voltage is
assumed to be the sum of two contributions. $V_{M}$ is associated
to the magnetic phase and $V_{S}$ to the superconductive (or
normal) phase.} \label{fig2}
\end{figure}
Additional two steps of annealing of 10~h in pure flowing argon at
1020~$^\circ$C also contributed to suppress the SrRuO$_3$ phase.
Subsequently, the powders were oxygenated. Seven oxygenation
cycles of a mean duration of 10~h, were performed at
1060~$^\circ$C in flowing pure oxygen. These fully oxygenated
powders were pressed in pellets by means of an hydrostatic press.
Five 10~h cycles in pure oxygen flux, at temperatures of
1050~$^\circ$C, 1055~$^\circ$C, 1060~$^\circ$C, 1065~$^\circ$C,
and 1070~$^\circ$C, with intermediate grinding and mixing, have
been performed on the pellets. Then, a last 90~h long cycle at
1070~$^\circ$C and a refining one of 10~h at 1065~$^\circ$C
assured the complete oxygenation of the pellets. The crystal
structure of the GdSr$_2$RuCu$_2$O$_8$ pellets was analyzed by
X-ray powder diffraction method. The data were collected with a
Philips PW-1700 powder diffractometer using Ni-filtered Cu
K$\alpha$ radiation. The X-ray spectrum of a typical fully
oxygenated pellet is shown in Fig.~\ref{fig1}.
 The scan pattern confirms that the sample is
GdSr$_2$RuCu$_2$O$_8$ single phased.

\section{Expected ac current-voltage curves}

If a magnetic phase is present in the
GdSr$_2$RuCu$_2$O$_8$, an hysteretic current-voltage ($I-V$) curve
should be expected when the current is swept with a frequency $\omega $.
\begin{figure}[tbp]
\includegraphics[width=7cm]{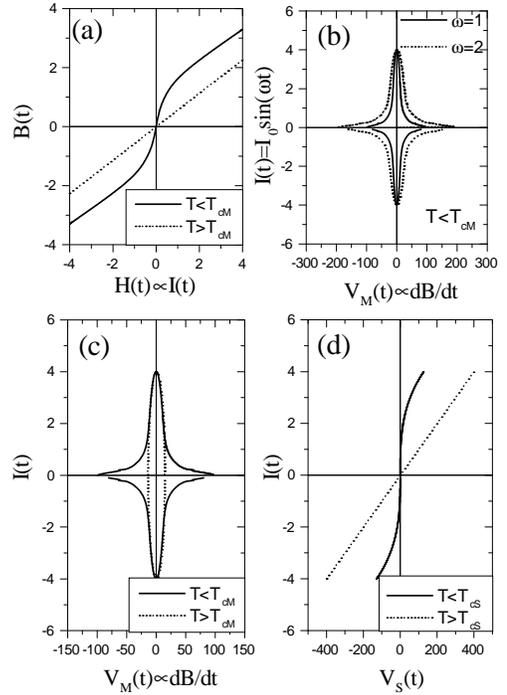}
\caption{(a) Low magnetic field approximation of $B(H)$ for the
ferromagnetic phase ($T<T_{cM}$) and the paramagnetic phase
($T>T_{cM}$). (b)$I-V_{M}$ curves of the ferromagnetic phase for
two different pulsations of a sinusoidal current forcing. (c) The
shape of ac $I-V_{M}$ curves is truly elliptical for the
paramagnetic phase and becomes a distorted ellipse in the
ferromagnetic phase. (d) Typical $I-V_{S}$ curves of the
superconductive ($T<T_{cS}$) or resistive ($T>T_{cS}$) phases.}
\label{fig3}
\end{figure}
In transport measurements, the four contact wire connection
sketched in Fig.~\ref{fig2}(a) is typically used. Here, the
forcing current $I(t)$ generates a magnetic field {\bf H}$(t)${\bf
=H}[$I(t)$] with an associated
magnetic induction {\bf %
B}$(t)${\bf =B[H}${\bf (}t{\bf )}${\bf ].} To the first order, the
magnetic field depends linearly on the forcing current,
$H(t)\propto I(t),$ so that $B(t)=B[I(t)]$ is too. From Maxwell
equations, we expect a voltage drop contribution due to the temporal
derivative of the magnetic flux linked to the voltage wires.
However, such a contribution is quite relevant only if the
magnetic induction field is quite high, i.e., if magnetic phases
are involved. The GdSr$_2$RuCu$_2$O$_8$ can be phenomenologically
seen as a series connection of superconducting and magnetic
phases. Hence, we expect the measured total voltage to be the sum
of a superconducting contribution $V_{S}$ and a magnetic
contribution $V_{M}$ [see Fig.~\ref{fig2}(b)]:
\[
\begin{array}{l}
V=V_{S}+V_{M} \\
V_{M}=\frac{d\Phi [B]}{dt}\propto
\frac{dB(t)}{dt}=\frac{dB[I(t)]}{dt}
\end{array}
\]
where we assumed that the relevant inductive voltage is
essentially due to the magnetic component. Some qualitative
predictions of the $I-V_{M}$ characteristic are possible from an
analysis of the expected $B[I(t)]$.

\begin{figure}[tbp]
\includegraphics[width=7cm]{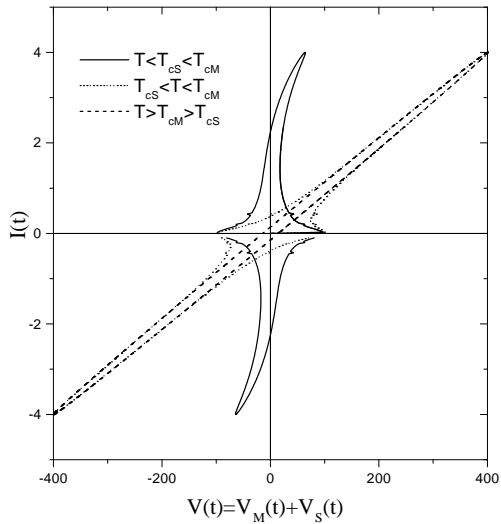}
\caption{ Total ac  current-voltage curve predicted for
GdSr$_2$RuCu$_2$O$_8$ pellet at three relevant
temperatures.} \label{fig4}
\end{figure}

Generally, {\bf B} is a nonlinear function of {\bf H } when a
material is in a magnetic phase. In the following we are concerned
with low magnetic field (forcing current) amplitudes. In such a
case, a linear relationship between {\bf B} and {\bf H} can be
assumed for the paramagnetic phase. Due to the vanishingly small
net magnetization, for an antiferromagnetic phase an
approximatively linear $B(H)$ relation could be again inferred,
while a nonlinear relation should be expected for a strongly
ordered phase as the ferromagnetic one. The last case can be
qualitatively discussed as follows. At a given low amplitude
magnetic field, a linear relation between {\bf B} and {\bf H} [see
Fig.~\ref{fig3}(a)] can be expected for temperatures above the
ferromagnetic transition temperature $T_{cM}$ (i.e., when the
material is in the paramagnetic phase) while a strongly nonlinear
relation between {\bf B} and {\bf H} should be  expected for
temperatures below $T_{cM}$ (i.e., when the material is in the
ferromagnetic phase). Due to the very low magnetic fields we can
generate with the normally used forcing currents (of the order of
some mA) we can assume that the saturation field will never
reached when the material is in the ferromagnetic phase. In other
words, for the used currents only the virgin curve of the
hysteresis loop will be normally swept, so that a single-valued
functional form $B(t)=B[I(t)]$ similar to the one shown in
Fig.~\ref{fig3}(a) can be expected to approximately describe the
material in the ferromagnetic phase. In such a limit, for a
sinusoidal forcing current of amplitude $I_{0}$ and pulsation
$\omega $  the $I-V_{M}$ curves shown in Fig.~\ref{fig3}(b) should
be observed for the ferromagnetic phase. Moreover, the distorted
ellipse typical of the ferromagnetic phase (at $T<T_{cM}$) should
become a pure ellipse in the paramagnetic phase (at $T>T_{cM}$),
as shown in Fig.~\ref{fig3}(c).

\begin{figure}[!h]
\includegraphics[width=7cm]{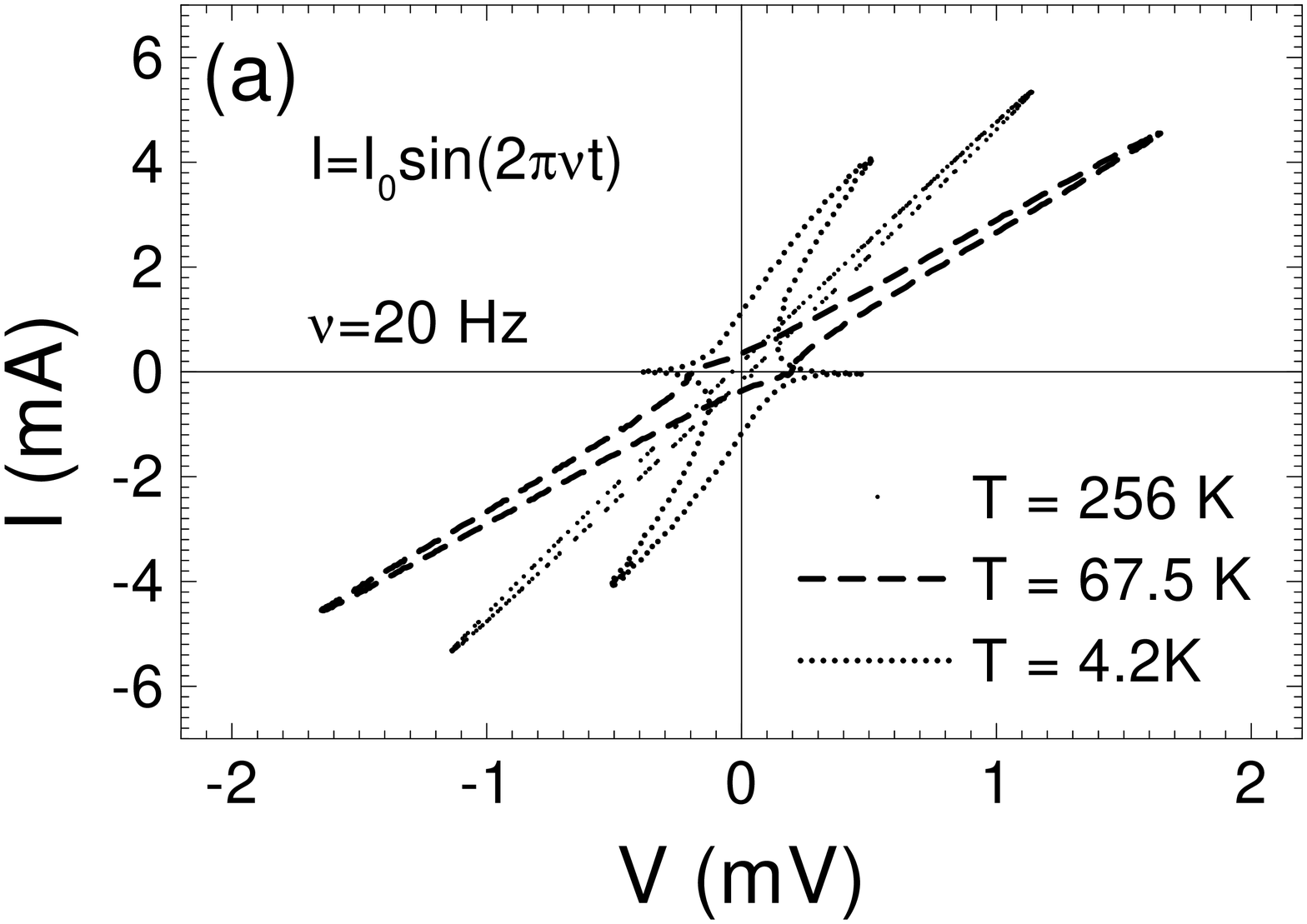}
\includegraphics[width=7cm]{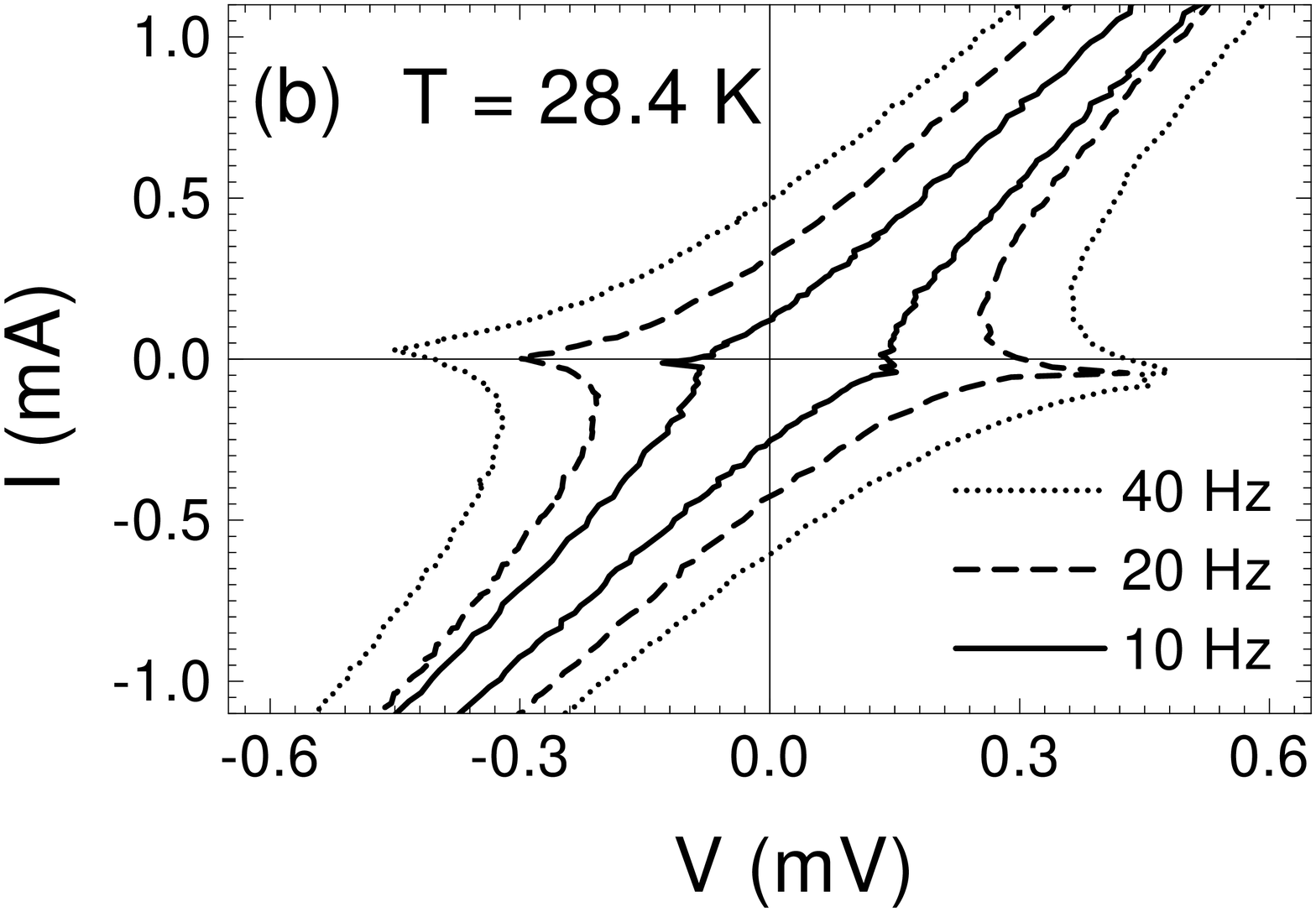}
 \caption{\label{fig5} (a) Experimental $I-V$ curves measured at three different temperatures.
 The frequency of current supply was 20 Hz.
  (b) $I-V$ curves measured at $T$=28.4~K at different frequencies}
 \end{figure}
\begin{figure}[!ht]
\includegraphics[width=7cm,clip]{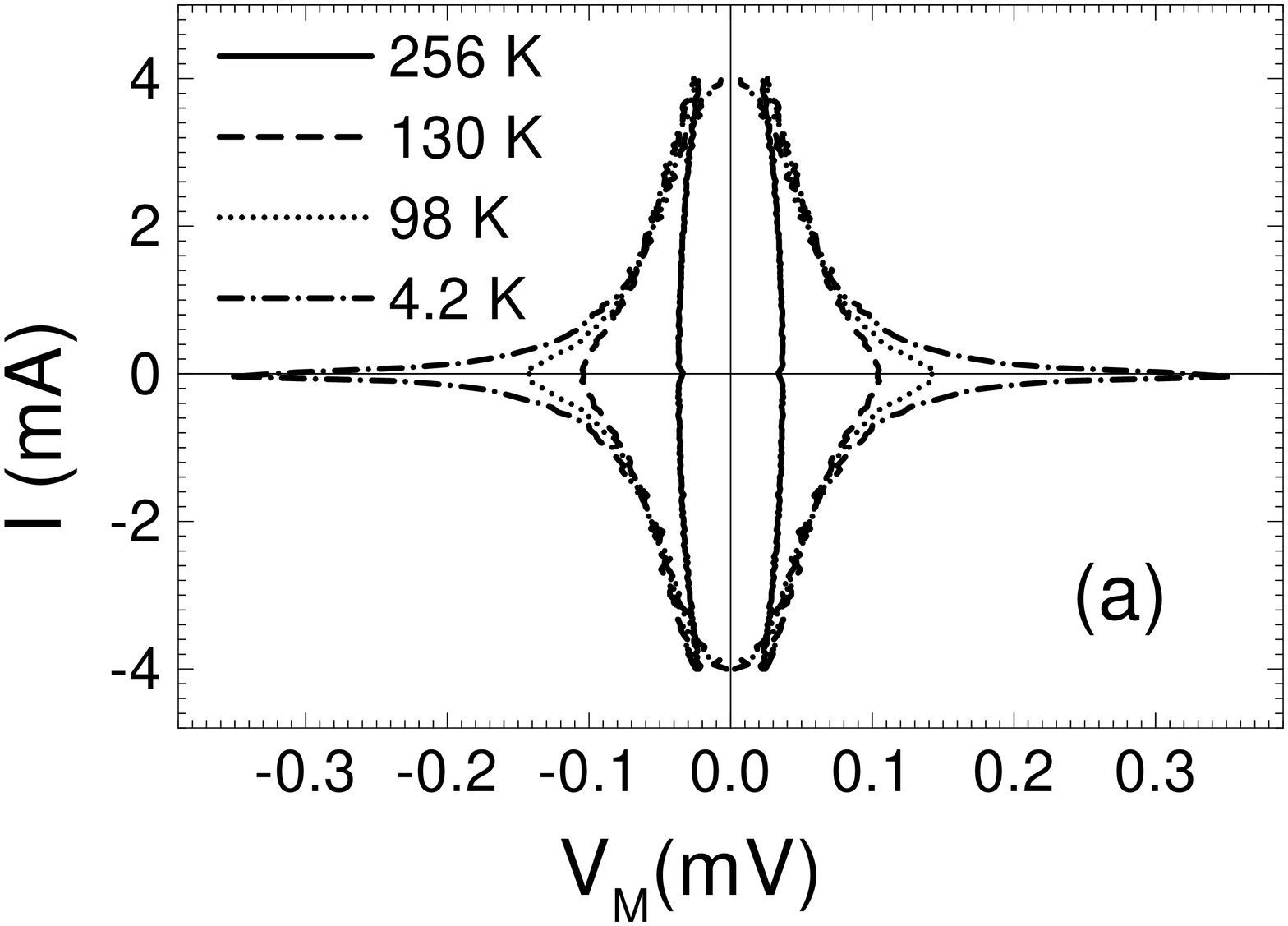}
\includegraphics[width=7cm,clip]{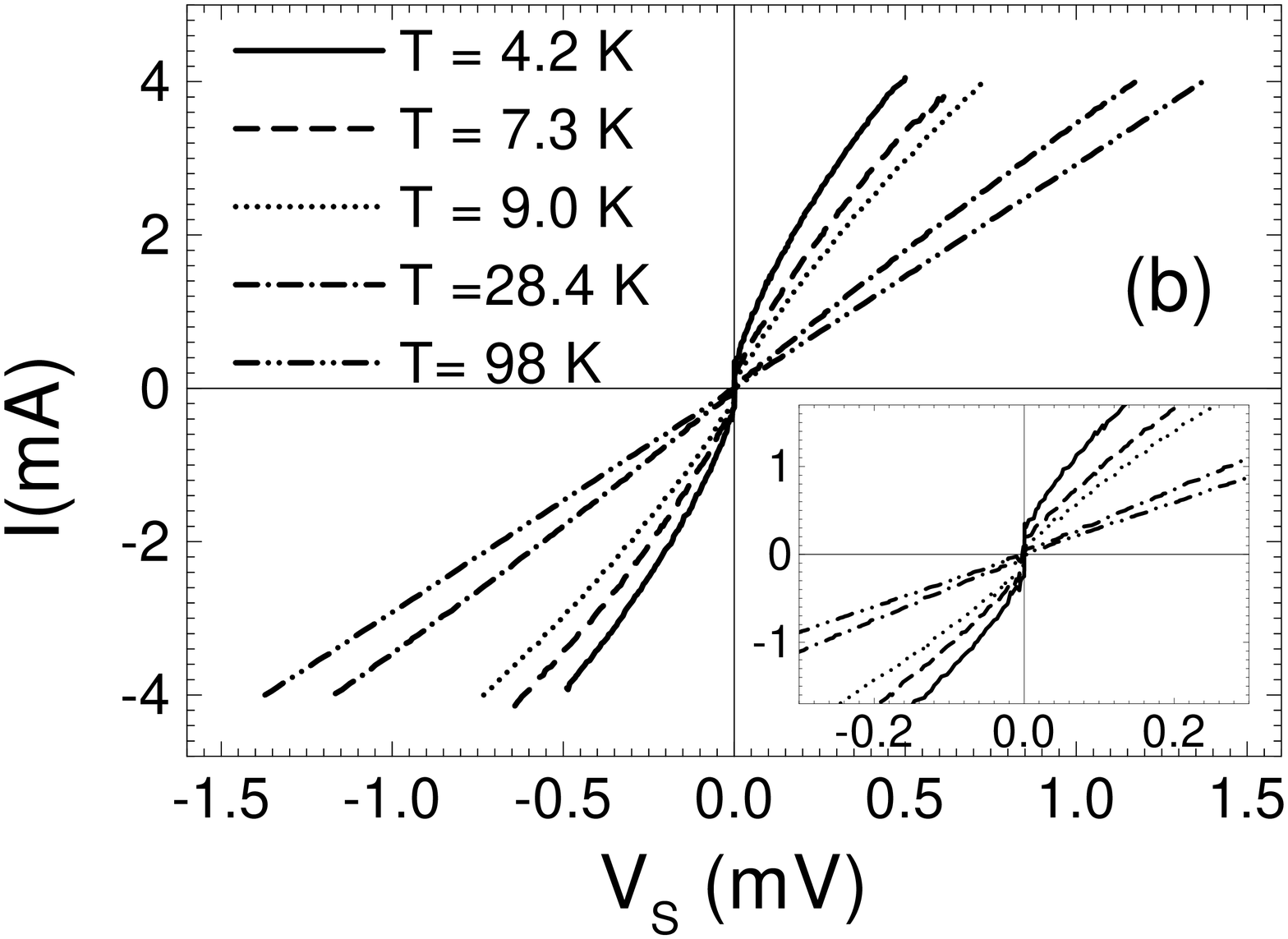}
 \caption{\label{fig6} Irreversible (a) and reversible (b) component
 extracted by experimental $I-V$ curves.}
 \end{figure}

Referring to the superconducting phase, the standard ac $I-V_{S}$
curves schematically plotted in Fig.~\ref{fig3}(d) are expected
for temperatures below or above the superconducting transition temperature $T_{cS}$. For $%
T<T_{cS}$ is $V_{S}=0$ for amplitude of the forcing current lower
than a critical current value $I_{c},$ while a truly resistive
curve is observed for $T>T_{cS}$

As stated above, the measured voltage of GdSr$_2$RuCu$_2$O$_8$
pellet is $V=V_{M}+V_{S}$. Hence, from information in
Figs.~\ref{fig3}(c) and (d), the expected ac $I-V$ curves should
look similar to the ones we plotted in Fig.~\ref{fig4} for three
relevant temperatures.

We should remark that, if observed, the peculiar outward cusp-like
distortion of the $I-V$ curve around $I=0$ is a signature of a
ferromagnetic order originating from the strong nonlinear increase
of the magnetic susceptibility below the Curie temperature.
Conversely, for an antiferromagnetic order, a smoother distortion
of the $I-V$ curve should be expected, and the area of the ellipse
of the paramagnetic phase should decrease for temperatures below
the Neel temperature due to the decrease of the
 magnetic susceptibility of the antiferromagnetic phase when
 the temperature is lowered.

\section{Transport measurements and discussion}

Measurements of $I-V$ curves were performed on a slice of
GdSr$_2$RuCu$_2$O$_8$ using the four contact technique shown in
Fig.~\ref{fig2}(a). The sizes of the slice were L=5~mm, W=2~mm,
and d=0.7~mm. A sinusoidal forcing current ranging from 4~mA to
6~mA and frequency values of 10, 20 and 40~Hz were used. In order
to reduce external electromagnetic interference, measurements were
performed in a shielded room. The sample was also enclosed in a
cryoperm shield to minimize external spurious magnetic field.

In Fig.~\ref{fig5}(a),
 $I-V$ curves recorded at three different
temperatures are shown. At first sight, the curves are in
qualitative agreement with the calculated ones [see
Fig.~\ref{fig4}], resulting from the phenomenological model
reported in the previous Section. The $I-V$ curves  show an
hysteretic behavior at each temperature measured. Below a certain
temperature, an outward cusp-like distortion of the elliptical
shape at T=256~K is evident in the curves. Moreover,  the loop
area always increases when temperature is lowered. As stated in
the previous section this means that a {\it ferromagnetic} phase
is involved in the material.

 Figure ~\ref{fig5}(b)
shows that the loop area increases with the frequency of the
current sweep, as expected for an inductive (magnetic)
contribution $V_{M}\propto dB/dt$ to the total voltage drop. By
comparison of  Fig.~\ref{fig5}(b) and Fig.~\ref{fig3}(b), a
voltage contribution from a ferromagnetic phase is achieved. In
the previous section, we have assumed that the electrical response
of the material can be described as the series connection of a
normal (superconducting/resistive) phase and a magnetic
(ferromagnetic/paramagnetic) phase. When the material is a.c.
supplied, the resistive (normal) component  gives a reversible
voltage signal, whereas the inductive (magnetic) one gives rise to
an irreversible response accounting for the hysteretic shape of
the voltage-current curves in the $I-V$ plane. In order to study
separately the resistive and the inductive components of the
measured $I-V$ curves, the reversible ($V_S$) and the irreversible
($V_M$) voltage were extracted  in each curve. The reversible
component in the total voltage signal, was calculated by using the
simple formula
\begin{equation}\label{eq1}
  V_S(I)=\frac{V_{up}(I)+V_{dw}(I)}{2}
\end{equation}
where $V_{up}$ and $V_{dw}$ are respectively the voltage values
measured during the increasing  and the decreasing branch of the
sinusoidal forcing current.  Then, the irreversible component was
extracted according to
\begin{equation}\label{eq2}
   V_M(I)=V(I)-V_S(I)
\end{equation}

The irreversible part extracted from  the total signal measured is
shown in Fig.~\ref{fig6}(a). Again, a qualitative agreement with
the computed curves [see Fig.~\ref{fig3}(b)] is recognized. For
temperatures ranging from 4.2~K up to about 70~K, the loop area
diminishes very slowly. Then the area decreases quickly and
smoothly changes shape becoming elliptical around $T_{cM}$=133~K.
From analysis of the previous Section we identify $T_{cM}$=133~K
as the Curie transition temperature of the magnetic phase in the
sample. In Fig.~\ref{fig6}(b) the reversible curves, ascribed to
the resistive share in the total voltage signal, are shown for
different temperatures. The typical non linear $I-V$ for
 a superconductor ($V_{S}=0$ for $-Ic<I<Ic$) can be recognized for
temperatures below $T_{cS}$=18~K while linear behavior is
recovered above this temperature.

\begin{figure}[!htb]
\includegraphics[width=7cm]{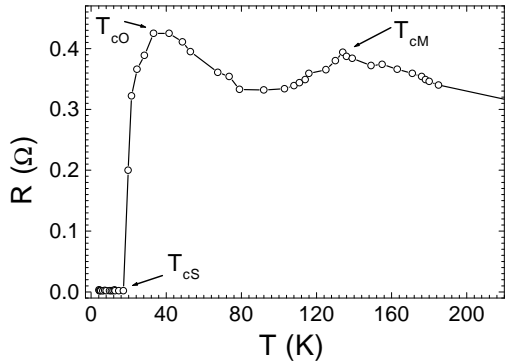}
\caption{\label{fig7} Resistance versus temperature calculated by
using the reversible $I-V$ curves.}
\end{figure}

From  data of the reversible curve, we extracted the resistance as
a function of the temperature shown in Fig.~\ref{fig7}. The
temperature $T_{cS}$, corresponding to a full superconducting
phase in the sample ($V_{S}=0$) and the onset temperature $T_{cO}$
were estimated 18~K and 33~K, respectively. In our measurements
the non-linear behavior in reversible $I-V$ curves, can be
recognized up to 18~K. Increasing the temperature from T$_{cS}$ up
to T$_{cO}$ the reversible $I-V$ curves are linear with a quite
fast increase of the resistivity. Above $T_{cO}$, the measured
resistance get lower and for $T_{cM}$=133~K the resistance shows a
peak. For temperatures above the magnetic transition temperature
the resistance diminishes again.

Two main results can be drawn from the data above presented.

Firstly, we find a clear evidence of changes near 130K of
irreversible and reversible components of the $I-V$ curves and
therefore we infer that they could be ascribed to a
ferromagnetic/paramagnetic-like transition. This speculation
agrees well with the results reported in literature that find a
magnetic ordering temperature at around 130K \cite{tallon}. The
appearance of a spontaneous magnetic moment below this temperature
at a very low field suggests that the transition at T$_{cM}$ must
have a significant ferromagnetic component. The experimental
results also suggest that the ferromagnetic component persists to
the lowest measured temperature attained in the experiments and
does not appear to weaken when the superconductivity comes in at
T$_{cO}$=33K. The existence of a ferromagnetic component in the
superconducting state of this sample suggested by the low
frequency data here presented, is also supported by magnetic
measurements performed on the same sample and reported elsewhere
\cite{eucas}. Moreover, because no impurity lines were detected in
the X-ray diffraction pattern within the experimental resolution
we may argue that no extra phases are responsible for
ferromagnetism implying that this ordering is due to an intrinsic
phase and in this respect we could infer that the coexistence of
superconductivity and ferromagnetism is realized within a
microscopic scale. This hypothesis is corroborated by
magneto-optical-imaging measurements where ferromagnetism and
superconductivity are directly observed to coexist in the same
space within the experimental resolution\cite{chu2}.

Second, we speculate briefly  on the significance of the
phenomenological model previously introduced. Within our model, we
assume that the measured total voltage is the sum of two
contributions: one coming from the superconducting channel and the
other  due to the  ferromagnetic ordering. Although the crudeness
of the assumptions, we have been able to reproduce fairly the
shape of the $I-V$ curves and more importantly we clearly identify
the superconducting contribution only when the ferromagnetic one
is subtracted. This contribution is of the standard form for a
generic superconductor and this in turn further supports the
correctness of our assumptions.
\section{Conclusion}

In conclusion, we performed measurements on
GdSr$_{2}$RuCu$_{2}$O$_{8}$ ruthenate-cuprate
with the aim to address the question of the
nature of the magnetic order in the superconducting phase and
trying to improve the understanding of the physics of
ruthenate-cuprate materials. We used a relatively
inexplored approach, based on the analysis of low
frequency electrical transport measurements.
The observed current-voltage curves
have been found in quite good qualitative agreement with
the predictions of a
phenomenological model
accounting for coexistence
of both magnetic and superconducting phases in the sample.
Our experimental results suggest that
GdSr$_{2}$RuCu$_{2}$O$_{8}$ is paramagnetic above
$T_{cM}$=133~K, ferromagnetic between
$T_{cS}$=18~K and $T_{cM}$=133~K, and
both ferromagnetic and
superconducting
 below $T_{cS}$=18~K.

\section{Acknowledgements}
We gratefully acknowledge the contribution of D. Sisti in the
sample preparation.

%\bibliography{lowfreq01}

\end{document}